\documentclass[twocolumn,aps,pra,amsmath,amssymb,showpacs,superscriptaddress]{revtex4}

\usepackage[utf8]{inputenc} 

\usepackage{graphicx}
\usepackage{graphics}
\usepackage{latexsym}
\usepackage{amsmath}
\usepackage{amsfonts}
\usepackage{amssymb}

\usepackage{hyperref}

\def\endproof{\vrule height6pt width6pt depth0pt}

\begin{document}

\title{
Wigner-PDC description of photon entanglement as a local-realistic theory}

\author{David Rodr{\'\i}guez}
\affiliation{
Departamento de F{\'\i}sica Aplicada III, Universidad de Sevilla, E-41092 Sevilla,
Spain}
\email{drodriguez@us.es}

\date{\today}


\begin{abstract}
\begin{center} 
\emph{(Revised, added references to recent tests)} 
\end{center}

The Wigner picture of Parametric Down Conversion
(works by Casado \textit{et al}) 
can be interpreted as a local-realistic formalism, without the need
to depart from quantum mechanical predictions at any step, at least for
the relevant subset of QED-states.
This involves reinterpreting the expressions for the detection probabilities, by
means of an additional mathematical manipulation;
such manipulation seemingly provides enough freedom to guarantee consistency
with the expectable, experimentally testable behavior of detectors, that being
nevertheless an unnecessary requirement in relation to our main results, of a
purely mathematical nature.
We also include an overview of the consequences of such results in
relation to typical Bell experiments.
%
\end{abstract}


\pacs{
03.65.Ud,
03.67.Mn,
42.25,
42.50.Xa
}
\maketitle

\noindent

\section{\label{PrevNote}
Previous notes on recent non-locality tests}

On the test of Giustina {\it et al}, 2015 
\cite{Giustina2015}:

This is a PDC-based experiment where the estimated "detection
efficiencies" are, very probably, not significant because they
are simply not well evaluated: 
the relevant parameter is a quotient, the denominator being a
detection rate in a particular setting, the numerator being a
proper average of the detection rates for each polarization at
the other arm, conditioned to one  obtained in the first... 
it seems that here the test only records or takes into account
one of those two last rates -just one detector after each
polarizer instead of the necessary two-;
rates of detection in just one of the polarizations may not be
(and I claim they will be proven not to be, if given the chance
to) statistically significant of what one would obtain if
both exit channels of the polarizer were detected, recorded:
that is the crucial point in this so-called "detection loophole"
(I use the inverted commas as the whole of the approach I
advocate here suggests that they are not substantially
diminisehd by any "detection efficiency",
but actually stand simply for the proper detection rates
predicted by an underlying wave-like theory).

And once here let me ask: 
why would we not once and for all estimate those rates in
the best possible way (with both detectors at each end)? 
As far as the information provided here goes, one cannot
say this does closes any "detection loophole".

On the tests of Hensen {\it et al}, 2015/2016 
\cite{Hensen2015,Hensen2016}:

How here is so readily stated that there is no "detection
loophole" remains an obscure, unsolvable mystery to me: 
regardless of the sophistication of the photon generation
method (customarily Parametric Down Conversion, here another
one based on a previous ``entanglement'' of the spins of
two massive particles), 
what we have at the end is, as usual, two light signals
travelling to two remote measurement settings
(polarizer-detector) and everything said in the paragraph
above applies exactly the same, taking this time into
account the rate limits for the inequality tested,
the CHSH one.

The first test has been followed by another one where some
technical aspects regarding other loopholes are polished: 
these other "loopholes" are not addresed here as in the
opinion of the author they are just that, mere "loopholes"; 
far beyond that, the detection loophole -and all that
is related to it, "enhancement" and so on- is 
clear evidence that the particle model with
which these tests are modeled may be useful on a first
approximation, but not subtle enough to describe the
real physics going on underneath, a physics that ultimately
excludes -how could it not?- all quantum states that are
not compatible with well defined, pre-existing probability
density functions for the results of all the measurements
that can be performed upon the system: local-realism.

And once again here, there is ONLY ONE WAY to be sure of what
is happening in these tests: to measure everything,
compute all the rates, exhaustively, not just a conveniently
chosen subensemble of the things one can observe in the
laboratory so that one or other Bell inequality -and/or its
additional assumptions- seems to be violated.

\section{\label{Intro}Introduction}

The Wigner picture of Quantum Optics of photon-entanglement
generated by Parametric Down Conversion \cite{PDC_explanation}
was developed already some decades ago in a series of
papers \cite{pdc1,pdc2,pdc3,pdc4,pdc5,pdc6,pdc7};
more recently, the approach has been revitalized producing
another stream of very interesting results
\cite{pdc8,pdc9,pdc_2014,pdc_2015,pdc_2018,pdc_2019}.

Starting from an stochastic electrodynamical description
(hence, based on continuous variables: electromagnetic fields
defined at each and every point of space),
by use of the so-called Wigner transformation \cite{KimNoz}
this model acquires a form where all expectation values
depend on a probability distribution (hence one objectively
defined) for the value of those fields in the vacuum.

Such a picture clearly differs from the one that is usually assumed in the field
of Quantum Information (QInf), based on a discrete description of light (particles)
and not on a wave-like model where photons could be perhaps be
understood as the abstraction exclusively related to the ``detection events'', 
indeed the only directly observable item for either the conventional (particle) model
or our alternative (wavelike) one here. 
In relation to this last, let us remark that we do not at any point need the
unprobable phenomenom of a ``wave-packet'' somehow managing to remain spatially
and temporarily confined: all we need is a distribution of the values of the
fields over the space-time continuun.
That, and the pressence of a random background -whether that is the quantum
electrodynamical one or another with adequate properties is indeed indifferent
to most of what we will say here-, a vacuum Zero-Point field (ZPF).
Those two elements, let us insist, are all we need: fields defined at every point 
of space, including a random, background component.

Once here, some differences between the two pictures become obvious even at
the most preliminary, qualitative analysys:
for instance, in this wave-like model the empty polarization channels
at either of the exits of a polarizing beam splitter (PBS) will become filled
with random components of that vacuum Zero-Point field (ZPF); 
such random components can for instance give rise to the enhancement of
the detection probability if the overall intensity arrived at the detector
is as a result increased: see note \cite{WignerPDC_enhancement}.
This comes up as a big difference with the customary model
of the set polarizer-detector is treated just as  a ``black box'', blocking
or just letting it go but always without the introduction of any additional
noise or ``information''.

It is then clear that the two models are not only internally different
but they can also give rise to different predictions, and in particular
that the Wigner-PDC framework offers considerably more ``room'' to accomodate
physical phenomena that its conventional counterpart:
for instance, the possibility of enhancement of the detection probability
at a polarizer, or simply the appearance of a variable probability at the
detector, dependent on the incoming intensity, something that one cannot
explain in the customary particle-like picture if it is not through the
inclusion of some external, artificially tuned ``detection efficiencies''
in the model -precisely our point that the low value of those efficiencies
should not be blamed on technological limitations alone.

The Wigner-PDC framework provides an alternative, local-realistic
explanation of the results of many typically quantum experiments, 
the result of a measurement depending on (and only on) the set of hidden
variables (HV) inside its light cone:
in this case, both the signal propagating from the source and the additional noise
introduced by the ZPF at intermediate devices and detectors.
Within a first series of published papers \cite{pdc1,pdc2,pdc3,pdc4,pdc5,pdc6,pdc7} 
those experiments included:
frustrated two photon creation via interference \cite{pdc1};
induced coherence and indistinguishableness in two-photon interference
\cite{pdc1,pdc2};
Rarity and Tapster's 1990 experiment with phase-momentum entanglement
\cite{pdc2};
Franson's (original, 1989) experiment \cite{pdc2};
quantum \textit{dispersion cancellation} and Kwiat, Steinberg and Chiao's
\textit{quantum eraser} \cite{pdc3}. 

From a most recent series 
\cite{pdc8,pdc9,pdc_2014,pdc_2015,pdc_2018,pdc_2019} 
we can add, 
on one side and amongst quantum cryptography experiments based on PDC,
two-qubit entanglement and cryptography \cite{pdc8} and quantum key
distribution and eavesdropping \cite{pdc8};
on the other, 
an interpretation for the experimental partial measurement of the Bell
states generated from a single degree of freedom (polarization) \cite{pdc9},
as well as polarization-momentum hyperentanglement \cite{pdc_2014},
entanglement swapping and teleportation \cite{pdc_2015}
and complete one-photon polarization-momentum Bell-state analysis 
\cite{pdc_2018};
most recently some more results have been obtained in relation to
optical quantum communication experiments \cite{pdc_2019}

As a "local realistic" theory, due to its the mathematical structure
based on well defined ``a priori'' probability density functions for the
local hidden values (the distribution of fields),
it is obvious that this formalism must give rise to predicted detection
rates  (the low values of which in some other context are  customarily
associated to a detection efficiencies),
are naturally bound to remain within the limits (what other call
``critical detection rates'') where it is already well acknowledged
that such an explanation may exist;
moreover, and this is a crucial point, those limits would appear as
just natural consequences of the theory: 
they are simply limiting values of detection rates 
-in particular conditional joint detection rates-
for states of light generated from a non-linear mix of a classical
signal (the laser) with a set of random components (the vacuum).
However,
that local-realistic interpretation was so far not devoid of other
difficulties, in particular related to the detection model (so far
merely a one-to-one counterpart with Glauber's original expressions 
\cite{detection}):
as a result of the normal order of operators, there average intensity
due to vacuum fluctuations is ``subtracted'', a subtraction that seems
to introduce problems related to the appearance of what could be
interpreted as ``negative probabilities''.

That last is the issue that interests us here, one that, as to be
expected, did motivate the proposal of several modifications upon
the expressions for the detection probabilities:
for instance, as early as in \cite{pdc4}, ``our theory is also in almost
perfect one-to-one correspondence with the standard Hilbert-space theory,
the only difference being the modification in the detection probability
that we proposed in relation...''.

Such modifications ranged from the mere inclusion of temporal  and
spatial integration \cite{pdc4} to the proposal of much more complicated
functional dependencies \cite{pdc7,W_LHV_02,MS2008},
all of them seeming to pose their own problems; for instance in
\cite{W_LHV_02},
a departure from the quantum predictions at low or high intensities,
experimentally disproved for instance in \cite{Brida_et_al02}.
Our route here is a different one however:
we do not propose any modification of the initial expressions for the
detection probabilities (in one-to-one correspondence with the initial
quantum electrodynamical model), but just explore the possibility of
performing some convenient mathematical
manipulation that casts them in a form consistent with the axioms of
probability, hence one consistent with local-realism.

Following for instance \cite{pdc4,pdc_2019} (see also \cite{detection}),
single detection probabilities can be expressed as
\begin{eqnarray}
P_{i}
&\propto& \langle I_i - I_{0,i} \rangle \nonumber\\
&=&  \int_{\alpha,\alpha^{*}}
( I_{i}(\alpha,\alpha^{*}) - I_{0,i} )\ W(\alpha,\alpha^{*})\ d\alpha d\alpha^{*},
\label{marginal}
\end{eqnarray}
and double -coincident- detections obey, at
least whenever there are conditions of space-like separation 
(see for instance \cite{pdc4,pdc_2019}),
the following expression
\begin{eqnarray}
P_{i,j} &\propto& 
\int_{\alpha,\alpha^{*}}
( I_{i}(\alpha,\alpha^{*})  - I_{0,i} ) \cdot ( I_{j}(\alpha,\alpha^{*})  - I_{0,j} )
\nonumber\\
&&\quad\quad\quad\quad\quad\quad\quad\quad\quad 
\times\ W(\alpha,\alpha^{*})\ d\alpha d\alpha^{*}, 
\label{joint}
\end{eqnarray}
where $\alpha,\alpha^{*}$ are vacuum amplitudes
of the (relevant set of) frequency modes \cite{note_modes} at the entrance
of the crystal, $W(\alpha,\alpha^{*})$ is obtained as the Wigner transform of the vacuum
state, $I_{i}(\alpha,\alpha^{*})$ is the field intensity (for that mode) and $I_{0,i}$ the
mean intensity due to the vacuum amplitudes, both at the entrance of the $i$-th detector 
(see \cite{note_modes}):
\begin{eqnarray}
I_0 = \int_{\alpha,\alpha^{*}}
I_{0}(\alpha,\alpha^{*})\ W(\alpha,\alpha^{*})\ d\alpha d\alpha^{*}.
\end{eqnarray}
The last three expressions would in principle allow us to identify the vacuum amplitudes
with a vector of hidden variables $\lambda \in \Lambda$ ($\Lambda$ is the space of events
or probabilistic space), with an associated density function $\rho(\lambda)$,
\begin{eqnarray}
\lambda       &\equiv& \alpha,\alpha^{*},    \label{lambda}\\
\rho(\lambda) &\equiv& W(\alpha,\alpha^{*}). \label{rho_lambda}
\end{eqnarray}
Now, for instance in \cite{pdc4} it is already acknowledged that (\ref{marginal})--(\ref{joint})
cannot, because of the possible negativity of the difference $I_{i}(\alpha,\alpha^{*}) - I_{0,i}$,
be written as
\begin{eqnarray}
P_{i}   & = & \int_{\Lambda}{ P_{i}(det|\lambda) \ \rho(\lambda)\ d\lambda }, \\
\label{lambda_single}
P_{i,j} & = &
\int_{\Lambda}{ P_{i,j}(det|\lambda)\ \rho(\lambda)\ d\lambda }, \nonumber\\
\label{lambda_joint}
\end{eqnarray}
where naturally $P_{i}(det|\lambda), P_{i,j}(det|\lambda)$ should stay positive (or
zero) always.
This last is our point of departure: in this paper we propose a reinterpretation of
the former marginal and joint detection probabilities based on a certain manipulation
of expressions (\ref{marginal})--(\ref{joint}).

The paper is organized as follows.
In Sec.\ref{No_pol} we will consider a setup with just a source and two detectors;
once that is understood, the interposition of other devices between the source and
the detectors poses no additional conceptual difficulty though it is nevertheless
convenient to address it in some detail: this will be done in Sec. \ref{Pol}.
The calculations in these two sections find support on the proofs provided in Appendix
\ref{Math}, and stand for our main result in this paper.
Sec. \ref{Fac} explores the question of ``$\alpha$-factorability'' in the model,
not only from the mathematical point of view but also providing some more physical
insights on its implications.

Up to that point the novel points of the paper are made and its results are
self-contained;
it is nevertheless natural to extend our analysis to some of their further
implications (amongst these, the consequences for Bell tests of local-realism,
and also a recent related proposal), in Sec. \ref{Comp},
where we also include a preliminary approach to questions regarding the physics of the
real detectors.
Finally, overall conclusions are presented in Sec. \ref{Conc}, and some
supplementary material is provided in Appendix \ref{Rev}, which may
not only help make the paper self-contained but perhaps also contribute to clarify
some of the questions addressed in this paper. 

\begin{figure}[ht!] \includegraphics[width=1.0 \columnwidth,clip]{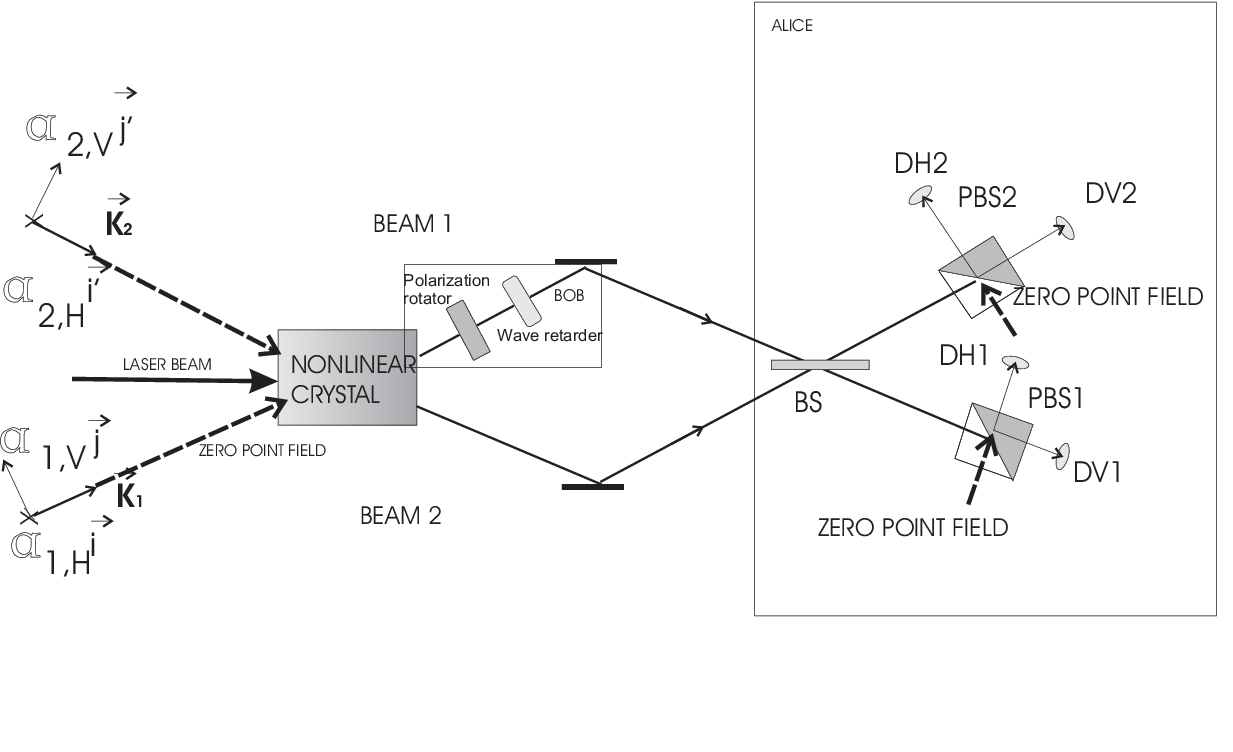}
\caption{
Wigner-PDC description for a typical experiment, including photon pair generation,
polarizing beam splitters (PBS) and detectors. 
Only relevant inputs of Zero-Point vacuum field (ZPF) are represented in the picture:
``relevant'' can be understood, in a classical wavelike approach, as ``necessary
to satisfy energy-momentum conservation'' for the  (set of) frequency modes of interest; 
in a purely quantum electrodynamical framework we would be talking about conservation
of the commutation relations at the empty exit channels of the devices. 
Besides, those new ZPF components introduced at the empty exit channels of PBS's 1
and 2 can alter the detection probability for the signal arriving to the detectors
(for instance giving rise to its ``enhancement'');
that signal is on the other hand determined (amongst other hidden variables) by the
ZPF components entering the crystal.
Figure: courtesy of A. Casado.
} \label{Scheme} \end{figure}


\section{\label{No_pol}
Reinterpreting detection probabilities}

Our aim is to adopt here an approach that is deliberately as abstract as it
can be chosen to be, because what we are concerned about is an (apparent)
problem of the mathematical structure of the theory, rather than other details
regarding its connection with physical reality which should in principle
be addressed at another stage of the investigation.
Nevertheless, of course for the sake of credibility some of these details 
need to be brought up, and so we will do when necessary.

We now propose a reinterpretation of expressions (\ref{marginal})--(\ref{joint}) 
which is based on the idea that it is legitimate to play with internal
degrees of freedom of a theory, as far as its observable predictions remain
all of them invariant to these transformations.
We intend to make a progressive exploration of the possible consistency
conditions, from less to more demanding, but we advance that in no way
this threatens the validity of our main result here, as we will be able
to provide proof of the existence of the required solution for all of them.

\subsection{Single detections}

Knowing that the following equality holds (from here on we drop detector indexes
when unnecessary), for some real constant $K_{(m)}$ (``marginal''),
\begin{eqnarray}
&&K_{(m)} \int_{\alpha,\alpha^{*}}
(I(\alpha,\alpha^{*}) - I_0)\ W(\alpha,\alpha^{*})\ d\alpha d\alpha^{*}
\nonumber\\
&&\quad\quad\quad\quad\quad\quad\quad\quad
= \int_{\Lambda}{ P(det|\lambda) \ \rho(\lambda)\ d\lambda },
\label{eq_marginal}
\end{eqnarray}
we realize we do not need to assume
\begin{eqnarray}
P(det|\lambda)  & \equiv & K_{(m)} \cdot (I(\alpha,\alpha^{*}) - I_0),
\end{eqnarray}
as a necessary, compulsory choice;
it would be enough to find some $f(x) \geq 0$, satisfying
\begin{eqnarray}
&& K_{(m)} \int_{\alpha,\alpha^{*}}
(I(\alpha,\alpha^{*}) - I_0)\ W(\alpha,\alpha^{*})\ d\alpha d\alpha^{*}
\nonumber\\
&&\quad\quad\quad\quad\quad\quad
= \int_{\alpha,\alpha^{*}}
f(I(\alpha,\alpha^{*}))\ W(\alpha,\alpha^{*})\ d\alpha d\alpha^{*},
\nonumber\\ \label{eq_marginal_1}
\end{eqnarray}
so we can then safely identify
\begin{eqnarray}
P(det|\lambda) & \equiv & f(I(\alpha,\alpha^{*})),
\end{eqnarray}
with $f(I(\alpha,\alpha^{*})) \geq 0$, $\forall I(\alpha,\alpha^{*})$.
This last is nothing but solving a linear system with only one restriction
and an infinite number of free parameters, whose subspace of solutions
intersects the region 
$0 \leq f(I(\alpha,\alpha^{*})) \leq 1$ $\forall \alpha,\alpha^{*}$:
see Appendix \ref{Math}.

\subsection{Joint detections}

From the perspective of our approach and regarding joint detections, it is not
very clear which are really the minimum conditions of consistence that one would
have to enforce;
in order to proceed with the maximum generality, let us simply define,
for detectors $i,j$ and a constant $K_{(j)}$ (``joint''), a new
function $\Gamma(x,y)$, so that
\begin{eqnarray}
&& K_{(j)} \int 
(I_i(\alpha,\alpha^{*}) - I_{0,i})\cdot(I_j(\alpha,\alpha^{*}) - I_{0,j})
\nonumber\\
&&\quad\quad\quad\quad\quad\quad\quad\quad\quad\quad\quad\quad\quad\quad
\times \ W(\alpha,\alpha^{*})\ d\alpha d\alpha^{*} \nonumber\\
&&\quad\quad
=
\int 
\Gamma(I_i(\alpha,\alpha^{*}),I_j(\alpha,\alpha^{*}))
\ W(\alpha,\alpha^{*})\ d\alpha d\alpha^{*}, \nonumber\\
\label{eq_joint_basic}
\end{eqnarray}
which we will now attempt to interpret as
\begin{eqnarray}
P_{i,j}(det|\lambda)
&\equiv& \Gamma(I_i(\alpha,\alpha^{*}),I_j(\alpha,\alpha^{*})). 
\end{eqnarray}
Of course, at a first look it seems very desirable to guarantee that
the detection probabilities depend solely on the amount of intensity
at the entrance of each detector; 
hence it would seem natural to add the condition
\begin{eqnarray}
\Gamma(I_i(\alpha,\alpha^{*}),I_j(\alpha,\alpha^{*}))
=
f(I_i(\alpha,\alpha^{*})) \cdot f(I_j(\alpha,\alpha^{*})), \nonumber\\
\label{eq_joint_alpha_fac}
\end{eqnarray}
i.e., some ``factorability'' on the incoming intensities.
\textit{Again, 
while such an additional condition seems necessary in order
to preserve the ``physical interpretability'' of the ``inner
structure'' of the theory, it is not at all something
necessary from the point of view of its observable predictions,
as long as these remain invariant;
as we will see later, $I_j(\alpha,\alpha^{*})$ may not only
be unobservable as corresponds to a particular realization
of the random fields... 
it may also happens that it does not actually represent a 
real intensity, but just an average promediated over another
relevant random variable.}

In Appendix \ref{Math} we have shown that it is always possible
to find some suitable $\Gamma$ satisfying all necessary conditions
to be interpreted as a probability distribution and consistent
with the observable prediction of the theory regarding joint detections,
eq. (\ref{joint});
and, furthermore, that a solution can be found satisfying also
(\ref{eq_joint_alpha_fac}).

As said, 
our choice here is to proceed without loss of generality: for
this purpose, it is convenient to redefine now
$\alpha \equiv \alpha,\alpha^{*}$,
as well as
\begin{eqnarray}
\hat{f}_i( \alpha )          &\equiv& f( I_{i}(\alpha)), \label{f_hat}\\
\hat{\Gamma}_{i,j}( \alpha ) &\equiv& \Gamma(I_i(\alpha),I_j(\alpha)). \label{Gamma_hat}
\end{eqnarray}
where we now assume that in general,
\begin{eqnarray}
&&\Gamma_{i,j}(\alpha) \neq  f_i(\alpha)\cdot f_j(\alpha),
\end{eqnarray}
The absence of factorability on the $\alpha$'s may come perhaps as a surprise
to some, given than Clauser-Horne factorability \cite{CH74}
on the hidden variable $\lambda$ is usually taken for granted;
this is a mistake \cite{factorability}, that we have tried to clarify
in Appendix \ref{Rev}.

\section{\label{Pol}
Adding intermediate devices:
polarizers, PBS's...}


Once we place one or more devices between the crystal and the detectors, 
typically  polarizers, polarizing beam splitters (PBS) or other devices to allow
polarization measurements (such as in Fig. \ref{Scheme}),
in general we cannot any longer describe the fields between both with only one
set $\{\alpha\}$ of mode-amplitudes;
we need to redefine our $\alpha$'s as now associated to a particular position
$\mathbf{r}$ (they do not any longer determine a frequency mode for all
space \cite{note_modes}).
Hence, we will now have
\begin{eqnarray}
\alpha(\mathbf{r}) \equiv
\{ \alpha_{\mathbf{k},\gamma}(\mathbf{r}), \alpha_{\mathbf{k},\gamma}^{*}(\mathbf{r}) \},
\end{eqnarray}
and, letting $\mathbf{r}_s$ be the position of the source (the crystal), and
$\mathbf{r}_i$ the position of the $i$-th polarizer or PBS (or any other
intermediate device), we will also redefine
\begin{eqnarray}
\alpha_s  &\equiv& \alpha(\mathbf{r}_s), \quad
\alpha_i   \equiv  \alpha(\mathbf{r}_i),
\end{eqnarray}
with $\alpha_i$ including the \emph{relevant} amplitudes at the (empty) exit
channels of that $i$-th intermediate device.
With $\alpha_s,\alpha_i$ corresponding each to (a set of) modes with different
(sets of relevant) wavevectors $\{\mathbf{k}_s\}$ and $\{\mathbf{k}_i\}$ \cite{note_wvectors},
we can then regard them as two (sets of) statistically independent random variables,
with all generality.
The intensity at the entrance of detector $i$-th will therefore depend now not only on
$\alpha_s$ but also on $\alpha_i$:
\begin{eqnarray}
P_{i}
&\propto& \langle I_i(\alpha_s,\alpha_i) - I_{0,i} \rangle \nonumber\\
&=&  \int_{\alpha_s} \int_{\alpha_i}
( I_{i}(\alpha_s,\alpha_i) - I_{0,i} )\ W(\alpha_s) \ W(\alpha_i)\ d\alpha_s d\alpha_i.
\nonumber\\ \label{marginal_p}
\end{eqnarray}
Once more, something like that can always be rewritten (see former section),
for some suitable and positively defined $f^{\prime}(x)$, as
\begin{eqnarray}
P_{i}
&=& \int_{\alpha_s} \int_{\alpha_i}
f^{\prime}( I_{i}(\alpha_s,\alpha_i))\ W(\alpha_s) \ W(\alpha_i)\ d\alpha_s d\alpha_i.
\nonumber\\
\end{eqnarray}
Integrating on $\alpha_i$ we would obtain
\begin{eqnarray}
P_{i} &=&  \int_{\alpha_s} \hat{f}^{\prime}_{i}( \alpha_s )\ W(\alpha_s) \ d\alpha_s,
\label{f_prime_hat}
\end{eqnarray}
from where we define a new function $\hat{f}^{\prime}_{i}( \alpha_s )$.
%
On the other hand, for joint detections we would have
\begin{eqnarray}
P_{i,j} &\propto&
\int_{\alpha_s} \int_{\alpha_i} \int_{\alpha_j}
( I_{i}(\alpha_s,\alpha_i)  - I_{0,i} ) \cdot ( I_{j}(\alpha_s,\alpha_j)  - I_{0,j} )
\nonumber\\ &&\quad\quad\quad\quad\quad
\times\ W(\alpha_s) \ W(\alpha_i)\ W(\alpha_j)\ d\alpha_s d\alpha_i d\alpha_j,
\nonumber\\ \label{joint_p}
\end{eqnarray}
which again can always be rewritten (again see former section), for
some positively defined $\Gamma^{\prime}(x,y)$, as
\begin{eqnarray}
P_{i,j} &=& \int_{\alpha_s} \int_{\alpha_i} \int_{\alpha_j}
\Gamma^{\prime}( I_{i}(\alpha_s,\alpha_i), I_{j}(\alpha_s,\alpha_j)) \nonumber\\
&&\quad\quad\quad\quad\quad
\times\ W(\alpha_s) \ W(\alpha_i)\ W(\alpha_j)\ d\alpha_s d\alpha_i d\alpha_j,
\nonumber\\ 
\end{eqnarray}
and integrating on $\alpha_i,\alpha_j$ we would obtain
\begin{eqnarray}
P_{i,j} &=& \int_{\alpha_s} \hat{\Gamma}^{\prime}_{i,j}( \alpha_s )\ W(\alpha_s) \ d\alpha_s,
\label{Gamma_prime_hat}
\end{eqnarray}
from where we can again define yet another new probability density
function $\hat{\Gamma}^{\prime}_{i,j}( \alpha_s )$.


It is interesting for the sake of clarity to compare the two situations: with (primed
functions) and without polarizers (unprimed).
It is easy to see that, because the detector only sees the intensity at its entrance
channel, clearly (we drop the ``s'' subscript for simplicity),
\begin{eqnarray}
f^{\prime}(x)        &=& f(x), \quad \forall x, \\
\Gamma^{\prime}(x,y) &=& \Gamma(x,y), \ \forall x,y.
\end{eqnarray}
while, in consistency with our approach, in general
$\hat{f}^{\prime}_{i}( \alpha ) \neq \hat{f}_{i}( \alpha )$,               
as well as
$\hat{\Gamma}^{\prime}_{i,j}( \alpha ) \neq \hat{\Gamma}_{i,j}( \alpha )$. 

\section{\label{Fac}On non-factorability}

\subsection{\label{Math_Fac}Mathematical analysis} 

Let us go back to the case with just the source and the detectors; we will soon
see the following does nevertheless also apply when polarizers or other devices
are added to the setup, just the same.                       
According to our reasonings in App. \ref{Rev}, and using (\ref{f_hat})--(\ref{Gamma_hat}),
we now realize that there is no way to avoid
\begin{eqnarray}
\hat{\Gamma}_{i,j}(\alpha) = \hat{f}_{i}(\alpha) \cdot \hat{f}_{j}(\alpha),
\end{eqnarray}
unless we introduce some additional dependence of the kind
$\hat{f}(\alpha) \rightarrow \hat{\hat{f}}(\alpha,\mu)$,
so that then
\begin{eqnarray}
\hat{\hat{\Gamma}}_{i,j}(\alpha,\mu)
\neq \hat{\hat{f}}_{i}(\alpha,\mu) \cdot \hat{\hat{f}}_{j}(\alpha,\mu),
\end{eqnarray}
where we add a second ``hat'' to avoid an abuse of notation, and where $\mu$
stands for a new set of random variables.
This $\hat{\hat{f}}(\alpha,\mu)$ should be interpreted as a detection
probability conditioned to the new vector of random variables $\mu$, i.e, 
\begin{eqnarray}
\hat{\hat{f}}(\alpha,\mu) \equiv P(det|\alpha,\mu).
\end{eqnarray}
We will impose further demands on $\hat{\hat{f}}(\alpha,\mu)$, defining
\begin{eqnarray}
\hat{\hat{f}}(\alpha,\mu) \equiv f(I(\alpha,\mu)),
\end{eqnarray}
something forced by strictly physical arguments: the choice $f(I(\alpha,\mu))$ must
prevail over other possible ones - for instance $f(I(\alpha),\mu)$ - due to the need
to respect the dependence of the probabilities of detection (conditioned to $\alpha$
or not) alone on the intensity that arrives to the detector, and nothing else.

Now, with the density function $\rho_{\mu}(\mu)$, we could write
\begin{eqnarray}
P(det|\alpha)       &=&  \int_{\mu} P(det|\alpha,\mu)\ \rho_{\mu}(\mu)\ d \mu, \\
P_{i,j}(det|\alpha) &=&  \int_{\mu} P_{i,j}(det|\alpha,\mu)\ \rho_{\mu}(\mu)\ d \mu ,
\end{eqnarray}
allowing us to recover our former definitions (\ref{f_hat})--(\ref{Gamma_hat}):
\begin{eqnarray}
\hat{f}(\alpha) \equiv P(det|\alpha), \\ 
\hat{\Gamma}_{i,j}(\alpha) \equiv P_{i,j}(det|\alpha).
\end{eqnarray}
For joint detections, the additional variable $\mu$ is particularly relevant
because, we will always have that while
\begin{eqnarray}
P_{i,j}(det|\alpha,\mu) =  P_{i}(det|\alpha,\mu) \cdot P_{j}(det|\alpha,\mu),
\end{eqnarray}
in general
\begin{eqnarray}
P_{i,j}(det|\alpha) \neq P_{i}(det|\alpha) \cdot P_{j}(det|\alpha),
\end{eqnarray}
or we could equivalently say that while necessarily
\begin{eqnarray}
\hat{\hat{\Gamma}}_{i,j}(\alpha,\mu) = \hat{\hat{f}}_{i}(\alpha,\mu) \cdot \hat{\hat{f}}_{j}(\alpha,\mu),
\end{eqnarray}
in general
\begin{eqnarray}
\hat{\Gamma}_{i,j}(\alpha) \neq \hat{f}_{i}(\alpha) \cdot \hat{f}_{j}(\alpha),
\end{eqnarray}
where of course
\begin{eqnarray}
\hat{\Gamma}_{i,j}(\alpha) = \int_{\mu}
\hat{\hat{f}}_{i}(\alpha,\mu) \cdot \hat{\hat{f}}_{j}(\alpha,\mu)\ \rho_{\mu}(\mu)\ d\mu.
\label{int_mu}
\end{eqnarray}

To conclude this section, we recover the case with intermediate devices:
due to $\alpha_i,\alpha_j$ being, as defined, independent from one another and also
from $\alpha_s$, our hypothetical ``flag'' $\mu$ cannot be associated with none of them.
Therefore, in general, and in principle, not only
\begin{eqnarray}
\hat{\Gamma}^{\prime}_{i,j}( \alpha ) \neq
\hat{f}^{\prime}_{i}( \alpha ) \cdot \hat{f}^{\prime}_{j}( \alpha ),
\label{Gamma_prime_nofac}
\end{eqnarray}
but also
$\hat{\Gamma}_{i,j}( \alpha ) \neq \hat{f}_{i}( \alpha ) \cdot \hat{f}_{j}( \alpha )$
either.
We use ``in principle'' because this question is not yet analyzed
in detail;
we now see clear, though, that this possible non-factorability on $\alpha$'s is
nothing more than an internal feature of the model's mathematical structure, 
bearing  no relevance in regard to its double-sided compatibility (or absence of
it) both with local-realism and the quantum predicted correlations.
A possible candidate for that additional hidden variable would be, at least in
my opinion, the phase of the laser $\mu$ \cite{laser_phase}.

\section{\label{Comp}Complementary questions}

\subsection{\label{LR_v_QM}
Wigner-PDC's local realism vs. quantum correlations}


Though former mathematical developments are fully meaningful and self-contained on
their own, yet it would be convenient to give some hints on how a local-realist (LR)
model can account for typically quantum correlations, which are known to defy that
very same local realism (LR).
In the first place and as a general answer, 
what the Wigner-PDC picture proves is that LR is respected by a certain subset of
all the possible quantum states, specifically the ones that can be generated from
a non-linear mix of the QED-vacuum (which therefore acts as an ``input'' for the model)
with a quasi-classical (a high-intensity coherent state), highly directional signal,
the laser ``pump'' (which indeed enters in the model as a non-quantized, external
potential \cite{pdc1}).
Moreover, such a restriction is clearly not arbitrary at all, since it arises from
a very simple quantum electrodynamical model of the process of generation of
polarization-entangled pairs of photons from Parametric Down Conversion (PDC):
see for instance eq. (4.2) in \cite{pdc1}.

\subsection{\label{Rates}
Detection rates and ``efficiencies''} 

Aside from subscripts, we will now also drop ``hats'' and ``primes'' for simplicity;
of course the fact that $\Gamma_{i,j}(\alpha)$ may not be in general $\alpha$-factorisable,
\begin{eqnarray}
\Gamma_{i,j}(\alpha) \neq f_i(\alpha) \cdot f_j(\alpha),
\end{eqnarray}
does not at all mean that it cannot well satisfy
\begin{eqnarray}
&&\int \Gamma_{i,j}(\alpha) \ W(\alpha)\ d\alpha = \nonumber\\
&&\quad\ \
\left[ \int f_i(\alpha)\ W(\alpha)\ d\alpha \right] \cdot
\left[ \int f_j(\alpha^{\prime})\ W(\alpha^{\prime})\ d\alpha^{\prime}
\right], \nonumber\\
\end{eqnarray}
i.e. (let us from now use superscripts ``W'' and ``exp'' to denote, respectively,
theoretical and experimental detection rates):
\begin{eqnarray}
P_{i,j}^{(W)}(det) =  P_{i}^{(W)}(det) \cdot P_{j}^{(W)}(det),
\end{eqnarray}
which is indeed the sense in which the hypothesis of ``error independence''
is introduced, to our knowledge, in every work on LHV models \cite{LHVs}.
This sort of conditions over ``average'' probabilities (average in the sense
that they are integrated in the hidden variable, may that be $\alpha$ alone
or also some other one) are the only ones that can be tested in the actual experiment; 
there, we can just rely on the number of counts registered on a certain
time-window $\Delta T$, and the corresponding estimates of the type
\begin{eqnarray}
P_{i}^{(exp)}(det) \approx \frac
{ n.\ joint \ det.\ (i,j) \ in\ \Delta T }
{ n.\ marg. \ det.\ (j)   \ in\ \Delta T}. \label{estimation_marg}
\end{eqnarray}
Now, if we wish to include some additional uncertainty element reflecting the
technological limitations (a ``detection efficiency'' parameter), what we have
to do is to redefine the overall detection probabilities as
\begin{eqnarray}
P_{i}^{(exp)}(det)   &\equiv& \hat{\eta}_i   \cdot P_{i}^{(W)}(det),   \label{eta_marginal}\\
P_{i,j}^{(exp)}(det) &\equiv& \hat{\eta}_i\hat{\eta}_j \cdot P_{i,j}^{(W)}(det), \label{eta_joint}
\end{eqnarray}
as well as
\begin{eqnarray}
P_{i}^{(exp)}(det|\alpha)   &\equiv& \hat{\eta}_i   \cdot P_{i}^{(W)}(det|\alpha),  
\label{eta_red_m} \\
P_{i,j}^{(exp)}(det|\alpha) &\equiv& \hat{\eta }_i\hat{\eta}_j \cdot P_{i,j}^{(W)}(det|\alpha),
\label{eta_red_j}
\end{eqnarray}
where $0 \leq \hat{\eta}_i \leq 1$ would play the role of such an (alleged) detection
efficiency, the ``hats'' remarking the fact that the customary definition of the
analogous quantity in QInf involves not only our $\hat{\eta}$'s but also the
non-technological contribution.
From the point of view of the experimenter it is very difficult to isolate both
components (Glauber's theory \cite{detection} does not predict a unit detection
probability even for high intensity signals);
we should perhaps then confine ourselves to the term ``observable detection rate''
instead of using the clearly misleading one of ``detector inefficiency''.

\subsection{
Consequences on Bell tests, their supplementary assumptions and critical efficiencies} 

That said and going to a lowest level of detail, states in such an (LR) subset
of QED can still indeed exhibit correlations of the class that is believed to collide
with LR, yet the procedure through which they are extracted from the experimental set
of data does not meet one of the basic assumptions required by every test of a Bell
inequality:
they do not keep statistical significance with respect to the physical set of
``states'' or hidden instructions \cite{LHVs}).
To guarantee that statistical significance we must introduce some of the following
two hypothesis:

\vspace{0.2cm}\noindent
(i) all coincidence detection probabilities are independent of the polarizers'
orientations $\phi_i,\phi_j$ (this is what we call ``fair-sampling'' \cite{fs},
for a test of an homogeneous inequality \cite{hom}), which implies
\begin{eqnarray}
P_{ij}(det|\phi_i, \phi_j, \alpha) = P_{ij}(det|\alpha_s), 
\end{eqnarray}
where of course (see Sec. \ref{Pol}) $\alpha \equiv \alpha_s \oplus \alpha_i \oplus \alpha_j$,
and where we recall that $\phi_i,\phi_j$ would determine which vacuum modes amongst the
sets $\alpha_i,\alpha_j$ would intervene in the detection process,\\

\vspace{0.2cm}\noindent
(ii) the interposition of an element between the source and the detector cannot
in any case enhance the probability of detection (the ``no-enhancement'' hypothesis
\cite{no_enhancement}, needed to test the Clauser-Horne inequality \cite{CH74}, and
presumably every other inhomogeneous one \cite{note_any}),
\begin{eqnarray}
P_{i}(det|\phi_i, \alpha) \leq P_{i}(det|\infty, \alpha_s).
\end{eqnarray}
with $\infty$ denoting the absence of polarizer and with
$\alpha \equiv \alpha_s \oplus \alpha_i$ this time.

\vspace{0.2cm}\noindent
Following our developments in Sec. \ref{Pol} one can easily see that (i) is not
in general true, and according to \cite{WignerPDC_enhancement} neither is (ii).
I.e., whenever states of light are prepared so as to produce the sort of quantum
correlations that are known to defy LR, these last come supplemented with the necessary
features that prevent it from happening... how could it not?


Yet, a mere breach of (i) or (ii) is not enough to assert the existence of a Local Hidden
Variables (LHV) model, which is an equivalent way of saying that the results of the
experiment respect LR: it is more than well known that this can only happen for certain
values of the observed detection rates \cite{LHVs,crit_eff}.
However,
from the point of view of (\ref{eta_red_m})--(\ref{eta_red_j}), and given the fact
that, as proven in Secs \ref{No_pol} and \ref{Pol}, the Wigner-PDC 
is in all circumstances in accordance with expressions (\ref{A_loc_causal})--(\ref{B_loc_causal}),
and hence to all possible Bell inequalities whether our $\hat{\eta}$'s are
equal or less than unity,
the so-called ``critical efficiencies'' would merely stand, at least as far as
PDC-generated photons are concerned, for bounds on the detection rates that we can
experimentally observe (these last in turn constrained by the only subset of quantum
states that we can physically prepare).


\subsection{\label{Sub}
An approach to realistic detectors and average ZPF subtraction}


This section approaches some physical considerations with the aim
of showing that there is plenty of room for a suitable physical interpretation of the
model, even when that is not strictly necessary for the coherence of our results,
at least from the purely mathematical point of view.
Indeed, we have already descended to the physical level when we established,
in former sections, the dependence of our $f$'s and $\Gamma$'s solely on the
intensity arriving at the detector.

The behavior we would expect from a physical device would typically include a
``dead-zone'', an approximately linear range and a ``saturation'' at high
intensities (this is indeed the kind of behavior suggested for instance in
\cite{W_LHV_02});
amongst other restrictions this would imply, for instance, $f(I(\alpha)) = 0$, when
$I(\alpha) \leq \bar{I}_0$ (this last a threshold that may even surpass the expectation
value of the ZPF intensity),
as well as $\Gamma(I_i(\alpha),I_j(\alpha)) = 0$ either for $I_i(\alpha)$ or $I_j(\alpha)$
below $\bar{I}_0$.

Neither these restrictions nor other similar ones would in principle invalidate our
proofs in Appendix \ref{Math}, which seem to provide room enough to simulate a wide
range of possible behaviors;
however, we must remark that none of our $f$'s and $\Gamma$'s can ultimately be
considered as fully physical models, due to the fact that they represent point-like
detectors
(the implications of such an over-simplification may become clearer in a
moment).

In close relation to the former, we also propose here a simple physical interpretation
of the term $-I_0$ appearing in the expressions for the detection probabilities.
From the mathematical point of view such subtraction arises from a mere manipulation of
Glauber's original expression \cite{detection}.
From the physical one, a realistic interpretation would be more than desirable, as that
subtraction of ZPF intensity is for instance crucial to explain the absence of an
observable contribution of the vacuum field on the detectors' rates \cite{non_detec_ZPF};
of course we mean ``explain in physical terms''; from the mathematical point of view
our model here already predicts a vanishing detection probability for the ZPF alone.

My suggestion is that the $-I_0$ term must be (at least) related to the average flux
of energy going through the surface of the detector in the opposite direction to the
signal (therefore \emph{leaving} the detector).
That interpretation fits the picture of a detector as a physical system producing
a signal that depends (with more or less proportionality on some range) on the
total energy (intensity times surface times time) that it accumulates.
I am just saying ``at least related'', bearing in mind that to establish
such association we would first have to refine the point-like model of a detector
which stems directly from the original Glauber's expression \cite{detection}.

\section{\label{Conc}Conclusions}


We have shown that the Wigner description of PDC-generated (Parametric Down Conversion)
photon-entanglement, so enthusiastically developed in the late nineties
\cite{pdc1,pdc2,pdc3,pdc4,pdc5,pdc6,pdc7} but then ignored in recent years,
can be reformulated as entirely local-realistic (LR).
A formalism that is one-to-one with a quantum (field-theoretical) model of the
experimental setup can be cast, thanks to an additional manipulation (also one-to-one),
into a form that respects all axiomatic laws of probability \cite{p_ax}, and therefore
LR, as defined for instance in Sec. \ref{Loc}.

The original quantum electrodynamical model takes as an input the vacuum state,
which accepts a well defined probabilistic description through the so-called Wigner
transform: this is the fact that the analogy with a local-realistic theory
is conditioned to.
What we call Wigner-PDC accounts, then, for a certain subset within the space of all
possible QED-states, determined by a particular set of initial conditions and a certain
Hamiltonian governing the time-evolution \cite{c_Ham},
restrictions that guarantee (according to our results here, but there are also 
certainly more intuitive ways of looking at it) the compatibility with LR.

Aiming for the maximum generality, as well as to avoid some possible (still under
examination) difficulties with factorisable expressions, we have renounced to what we
call $\alpha$-\emph{factorability} of the joint detection expressions.
Such a choice is not only perfectly legitimate \cite{factorability}, but may also be
supported by a well feasible interpretation (see Appendix \ref{Fac});
nevertheless, further implications of that non-factorability on $\alpha$ will also be
left to be examined elsewhere.
Again, whatever they finally turn out to be, they are also irrelevant for our main
result in this paper:
the Wigner-PDC formalism can be cast into  a form that respects all axiomatic laws of
probability for space-like separated events.

Neither does the explicit distinction between the cases where polarizers (or other devices
placed between the source and the detectors) are or not included in the setup introduce
any conceptual difference from the point of view of our main result;
however, the question opens room to remark some of the main differences of the Wigner-PDC
model (actually, also its Hilbert-space analogue) with the customary description used
in the field of Quantum Information (QInf):
here, each new device introduces noise, new vacuum amplitudes that fill each of its
empty polarization channels at each of its exists, in contrast to the usual QInf ``black
box'', able to extract polarization information from a photon without any indeterministic
component.

As a matter of fact, those additional random components may hold the key to explain the
variability of the detection probability (see note \cite{WignerPDC_enhancement}) that
is necessary, from the point of view of Bell inequalities, to reconcile quantum
predictions and LR (see Sec. \ref{LR_v_QM}).
In particular, the immediacy with which the phenomenon of ``detection probability
enhancement'' arises in the Wigner-PDC framework would suggest that this may be after
all the right track to understand why after several decades the minimum
detection rates (or, in QInf terminology, critical detection efficiencies) that
would lead to obtain conclusive evidence of non-locality are still out of reach
(see an explanation of what these critical rates are in \cite{crit_eff}, for comments
on the current state of the question see somewhere else).

In Sec. \ref{Sub} we have done a first, general approach on the question of whether
our reinterpretation of the detection probabilities is consistent with the actual
physical behavior of detectors.
A closer look to this question is out of the scope of the paper;
former proposals \cite{W_LHV_02} in regard to this issue aimed perhaps too
straightforwardly to the physical level, while they did not even guarantee consistency
with the framework we have settled here 
(proof of this is that it introduces divergences in relation to the purely quantum
predictions, divergences later experimentally disproved for instance in
\cite{Brida_et_al02}).
To conclude, we have also suggested a possible simple
physical interpretation of $I_0$-subtraction taking place in the expressions
for the detection probabilities:
work in any of these directions would anyway seem to require a departure from
the point-like model of a detector.

In summary, we have shown that a whole family of detection models
\begin{eqnarray}
{\cal M}_{det} \equiv \{f(I(\alpha,\mu)),\Gamma(I_i(\alpha,\mu),I_j(\alpha,\mu)\},
\end{eqnarray}
can be found, consistent both with the quantum mechanical expressions from the
Wigner-PDC model and LR.

A close examination of the constraints coming from the physical behavior of the detectors
and other experimentally testable features is left as a necessary step for the future,
with the aim of establishing a subset 
physically feasible ones;
nevertheless, we have the guarantee that all of them produce suitable predictions,
as so does their quantum electrodynamical counterpart.

Yet, even at the purely theoretical level some other features remain open
too: as a fundamental one, to what extent the model
requires what we have called non-factorability on $\alpha$'s.

Another limitation of this work that should be remarked is that (\ref{joint}),
taken here for the joint detections, is only valid \cite{pdc4,pdc_2019} 
when space-like separation is  guaranteed, a more general expression 
\cite{detection} being needed when it is not so; 
the compatibility of this last with our program here shall be analyzed somewhere
else, such nevertheless standing for little a drawback on the present work
since space-like separation is a necessary requirement of most of Bell tests.

Once more and after all, 
QM is just a theory, a theory that provides a formalism upon which to build
models, models than can (and should) be refined based on experimental evidence,
something that (again) also applies just as well to the one we are dealing with
here \cite{note_WPDC}.
Perhaps the particle properties of light are not enough to assume that the
current model of ``a photon'' is the best representation (and most complete)
that we can achieve of light;
many of those properties can be understood, I believe, from entirely classical
models \cite{classic_photon}, as well.
Quantum entanglement seems to manifest in many of its ``reasonable'' features but
at least as (this particular model of) Parametric Down Conversion is concerned
and so far, whenever local-realism would seem to be challenged new phenomena can
be invoked so as to prevent, at least potentially, that possibility.
Such are ``unfair sampling'', ``enhancement'' (as a particular case of variable
detection probability) and over all detection rates low enough to open room for
the former two.

Those phenomena find theoretical support in the Wigner-PDC picture, but
definitely not in the usual, based merely on the correspondence
between the $\tfrac{1}{2}$-spin algebra for massive particles and the
polarization states of a plane wave
(a photon then looks just as a $\tfrac{1}{2}$-spin particle, what some like to 
call a two-level system, but for the magnitude of the angular momentum it
carries, and its statistics of course).
To explore, and exhaust if that is the case, alternative routes such as the one
here is not only sensible but also necessary.

\section*{Acknowledgments}

I thank R. Risco-Delgado, J. Mart{\'i}nez and A. Casado for very useful discussions,
though most of them at a much earlier stage of the manuscript.
They may or may not agree with either the whole or some part of my conclusions and
results, of course.

\appendix

\section*{\label{Ap}Appendix}

\subsection{\label{Math}Auxiliary proofs}

\vspace{0.2cm}\noindent\emph{Lemma 1}:
\emph{There always exists $f(x)$ satisfying (\ref{eq_marginal_1}), with
$0 \leq f(x) \leq 1$, for $x=I(\alpha,\alpha^{*})$ over the space of all
possible pairs $\{\alpha,\alpha^{*}\}$.}

\vspace{0.2cm}\noindent\emph{Proof:}
We have a linear system with one restriction and infinite variables:
the set of values $\{f(I(\alpha,\alpha^{*}))\}$;
coefficients are given by the $W(\alpha,\alpha^{*})$'s, and the independent
term is the left-side term of (\ref{eq_marginal_1}), $0 \leq P_{s} \leq 1$.

Then let us consider the real, vectorial space ${\cal V}$ where we associate
each value $f(I(\alpha,\alpha^{*}))$ per coordinate: 
we can assume $\dim({\cal V}) = N^2$ ($N \rightarrow \infty$) if we consider
a direct dependence on the $\alpha$'s, 
or we can also say $\dim({\cal V}) = N$ (again $N \rightarrow \infty$) if the
dependence is defined upon the intensities, in principle a real number
(depends on whether we choose to formulate the model upon instantaneous values,
therefore real, or as a complex amplitude, this is not essential here).
Both options, $f$'s depending either on $\{\alpha\}$ or intensities 
$\{I(\alpha,\alpha^{*})\}$ serve right for our purposes, 
the second being perhaps more appropriate from the point of view of
physical interpretation.
If they do exist, compatible solutions $f(I(\alpha,\alpha^{*}))$ for the (unique)
restriction will conform a linear manifold ${\cal M} \subset {\cal V}$.
Now we have to see:

(i) Due to $W(\alpha,\alpha^{*}) > 0$ for all pairs $\alpha,\alpha^{*}$
\cite{W_expression}, and $P_{s} \geq 0$ too, ${\cal M}$ cannot be parallel with
any of the coordinate hyper-planes in ${\cal V}$; i.e., ${\cal M}$ intersects all
of them, defined each (each one for a pair $\alpha,\alpha^{*}$) by the equation
$f(I(\alpha,\alpha^{*})) = 0$.

(ii) Moreover, for $P_{s} > 0$, ${\cal M}$ has a non-trivial intersection
(more than one point, the origin) with all coordinate planes inside the first
hyper-quadrant ${\cal V}_{1q}$ (the subregion ${\cal V}_{1q} \subset {\cal V}$
given by restricting ${\cal V}$ to $f(I(\alpha,\alpha^{*})) \geq 0$).
This can be seen, for instance, determining the point of crossing
with the axes: doing all $f$'s zero except one (always possible because all the
$W$'s are strictly above zero), we can see the crossings always take place
at the positive half of the corresponding coordinate axis, therefore at the
boundary of ${\cal V}_{1q}$.
On the other hand, for $P_{s} = 0$ the solution for the system is trivial.

With (i) and (ii), it is clear that under the restriction (under the set of
inequalities) $f(I(\alpha,\alpha^{*})) \leq 1$ $\forall \alpha,\alpha^{*}$, the
set of admissible solutions is still not empty, as guaranteed by
\begin{eqnarray}
\int_{\alpha,\alpha^{*}} W(\alpha,\alpha^{*})\ d\alpha d\alpha^{*} = 1,
\label{W_norm}
\end{eqnarray}
and $P_{s} \leq 1$, as we will now show.
Here we give an inductive reasoning;
let us consider the equivalent problem in $3$ dimensions, with
\begin{eqnarray}
a x + b y + c z = d;
\end{eqnarray}
clearly, for $a,b,c,d \geq 0$ the former plane always intersects
with $x=y=z$ at a point of coordinates
\begin{eqnarray}
x = y = z = d/(a+b+c) = d,
\end{eqnarray}
which is obviously in the first quadrant ($x,y,z \geq 0$), and moreover,
for $a + b + c = 1$ and $0 \leq d \leq 1$, clearly also inside of the region
$\{ 0 \leq x \leq 1, 0 \leq y \leq 1, 0 \leq z \leq 1 \}$.
The extension to infinite dimensions, topological abnormalities all absent, 
is direct, with $x \equiv \alpha,\alpha^{*}$ (or alternatively
$x \equiv I(\alpha,\alpha^{*}$), $a,b,c \equiv W's$ and $d \equiv P_{s} \leq 1$, 
the left hand side of (\ref{eq_marginal_1}).
\hfill\endproof

\vspace{0.2cm}\noindent\emph{Lemma 2}:
\emph{There always exists $\Gamma(x,y)$ satisfying (\ref{eq_joint_basic}), with
$0 \leq \Gamma(x,y) \leq 1$, for $x=I_i(\alpha,\alpha^{*})$, $y=I_j(\alpha,\alpha^{*})$
over the space of $\{\alpha,\alpha^{*}\}$.}

\vspace{0.2cm}\noindent\emph{Proof:} formally identical with Lemma 1.
\hfill\endproof

\vspace{0.2cm}\noindent\emph{Lemma 3}:
\emph{There always exists $f(x)$ satisfying simultaneously (\ref{eq_marginal_1})
for a finite collection of $i=1, \ldots,N$ detectors, with
$0 \leq f(x) \leq 1$, for $x=I_i(\alpha,\alpha^{*})$, and satisfying, for any two
pair of detectors, condition (\ref{eq_joint_alpha_fac}) as well.}

\vspace{0.2cm}\noindent\emph{Proof:} 
First we notice that condition (\ref{eq_marginal_1}) for each additional
detector stands for a new linear restriction on the problem already treated
in \textit{Lemma 1};
due to the system being under-determined with an infinite number of
free parameters this poses no problem.

However,
condition (\ref{eq_joint_alpha_fac}) is not a linear but a non-linear
restriction.
Consider first the case $N=2$ (number of detectors), and let us solve the
system under the $2$ restrictions (one per detector) of the kind
(\ref{eq_marginal_1}), obtaining a solution in accordance with
\textit{Lemma 1};
such solution, as seen, depends on an infinite set of free parameters. 
Let us give values to all these free parameters (from Lemma 1,such
values can be chosen so as to guarantee $0 \leq f(x) \leq 1, \forall x$)
but one, which we will define as
\begin{eqnarray}
f(I_{\Theta}) = \Theta,
\end{eqnarray}
for some particular (randomly chosen) value $I(\alpha) = I_{\Theta}$ 
(the same value of $I(\alpha)$ can be produced by several $\alpha$'s,
so it is clearer to associate ``labels'' to $I$ instead of $\alpha$,
though conceptually botch choices are equivalent);
condition (\ref{eq_joint_alpha_fac}) between the two detectors stands
now for a quadratic equation,
\begin{eqnarray}
A \cdot \Theta^2 + B(I_\Theta) \cdot \Theta + C(I_\Theta) = D;	
\label{eq2}
\end{eqnarray}
before we make the former terms explicit, let $\Omega$ be the set
configured by all possible values of $\alpha$, and define the subsets
\begin{eqnarray}
\hat{\bar{\Omega}}(I_\Theta) 
&\equiv&  
\{\alpha; I_1(\alpha) = I_2(\alpha) = I_{\Theta}\}, \\
\bar{\Omega}(I_\Theta) 
&\equiv& 
\{\alpha; I_1(\alpha) \neq I_{\Theta}, I_2(\alpha) \neq I_{\Theta}\}, \\
\hat{\Omega}(I_\Theta) 
&\equiv&
\Omega - \hat{\bar{\Omega}}(I_\Theta) - \bar{\Omega}(I_\Theta),
\end{eqnarray}
so that now we can write
\begin{eqnarray}
A(I_\Theta) 
&=& 
\int_{\hat{\bar{\Omega}}(I_\Theta)}  d\alpha \ W(\alpha), \\
B(I_\Theta) 
&=& 
\int_{\hat{\Omega}(I_\Theta)} 	
d\alpha \ W(\alpha) \cdot
\left[f(I_1(\alpha)) + f(I_2(\alpha))\right], 
\nonumber\\
\\
C(I_\Theta) 
&=& 
\int_{\bar{\Omega}(I_\Theta)} 	
d\alpha \ W(\alpha) \cdot f(I_i(\alpha) \cdot f(I_j(\alpha)).
\end{eqnarray}
Taking now into account that intensities are continuous functions
of $\alpha$ it is reasonable to assume
\begin{eqnarray}
A(I_\Theta) \approx 0; 
\end{eqnarray}
and now with $0 \leq B,C,D \leq 1$ and clearly $C \leq D$ (just
look at eq.\ref{eq2}),
and finally the equivalence of $B$ to a mere single detection
probability (eq. \ref{eq_marginal_1} for instance),
we can now write
\begin{eqnarray}
B(I_\Theta) \geq D \geq D - C(I_\Theta) \geq 0,
\end{eqnarray}
from where it is not difficult to see that eq.(\ref{eq2})
always admits a solution $0 \leq \Theta \leq 1$, where we
recall the definition of the free parameter as
$\Theta \equiv f(I_\Theta)$.

Extension to $N$ detectors is inmediate by considering $N-1$
free parameters and repeating the former reasoning on each one
of them at a time, for each corresponding pair of detectors.
\hfill\endproof

\vspace{0.2cm}\noindent
Lemmas 1,2 and 3 are directly applicable, aside from Sec. \ref{No_pol},
to Sec. \ref{Pol}.

\subsection{\label{Rev}Basic concepts revisited}



We first briefly revisit the concepts of locality, determinism and factorability;
a good understanding on these concepts is crucial for the main results of the paper,
what makes this review not only convenient but almost unavoidable, especially given
the presence of some confusion in the literature.
In any case, it shall be clearly understood that \textit{Clauser and Horne's
factorability \cite{CH74} is not a requisite for local-realism. }

\subsubsection{\label{Loc}Locality and realism}

A theory predicting the results of two measurements $A$ and $B$ that take place
under causal disconnection (relativistic space-like separation) can be defined as
local if and only if we can write
\begin{eqnarray}
A = A(\lambda,\phi_A), \quad B = B(\lambda,\phi_B), \label{loc_det}
\end{eqnarray}
where $\lambda$ is a (set of) hidden (or explicit) variables defined inside the intersection
of both light cones,
and $\phi_A,\phi_B$ are another two other sets of variables (amongst them the configurable
parameters of the measuring devices) defined locally at $A$ and $B$, respectively, and
causally disconnected from each other, i.e.,
\begin{eqnarray}
P(A = a | \lambda,\phi_A,\phi_B ) = P(A = a | \lambda,\phi_A ), \label{A_loc_causal}\\
P(B = b | \lambda,\phi_A,\phi_B ) = P(B = b | \lambda,\phi_B ). \label{B_loc_causal}
\end{eqnarray}
These last two expressions are usually taken as a definition of \textit{local causality}
\cite{local_causality}.

Now, realism simply stands for $\lambda$ (and $\phi_A,\phi_B$ as well) having
a well defined probability distribution.
A set of physical observables corresponding to a particular quantum state can
be sometimes described by a well defined joint probability density (such is
the case of field amplitudes in any point of space for the vacuum state in QED);
for other quantum states that is not possible though.


\subsubsection{\label{Det}Determinism}

A measurement $M$ upon a certain physical system, with $k$ possible outcomes
$m_k$, is \emph{deterministic} on a hidden variable (HV) $\lambda$ (summarizing
the state of that system), if (and only if)
\begin{eqnarray}
P(M = m_k |\lambda) \in \{0,1\}, \ \forall k, \lambda, \label{det}
\end{eqnarray}
which allows us to write 
\begin{eqnarray}
M \equiv M(\lambda),
\end{eqnarray}
and \emph{indeterministic} iff, for some $\lambda$, some $k^{\prime}$,
\begin{eqnarray}
P(M = m_{k^{\prime}} |\lambda) \neq \{0,1\}, \label{indet}
\end{eqnarray}
i.e., at least for some (at least two) of the results for at least
one (physically meaningful) value of $\lambda$.

Now, indeterminism can be turn into determinism, i.e., (\ref{indet})
can into (\ref{det}), by defining a new hidden variable $\mu$, so that
now, with $\gamma \equiv \lambda \oplus \mu$:
\begin{eqnarray}
P(M = m_k |\gamma) \in \{0,1\}, \ \forall k,\gamma, \label{indet_ng}
\end{eqnarray}
which means we can write, 
\begin{eqnarray}
M \equiv M(\lambda,\mu),
\end{eqnarray}
a proof that such a new hidden variable $\mu$ can always be found (or
built) given in \cite{proof_1}.

So far, then, our \emph{determinism} and \emph{indeterminism} are
conceptually equivalent, though of course they may correspond to different
physical situations, depending for instance on whether $\gamma$ is
experimentally accessible or not.

\subsubsection{\label{Fac_concept}Factorability}

Let now ${\cal M} = \{M_i\}$ be a set of possible measurements, each
with a set $\{m_{i,k}\}$ of possible outcomes, not necessarily isolated
from each other by a space-like interval.
We will introduce the following distinction: we will say ${\cal M}$ is

\vspace{0.2cm}\noindent
(a)
\emph{$\lambda$-factorisable}, iff
we can find a set $\{\xi_i\}$ of random variables, \emph{independent
from each other and from $\lambda$ too}, such that
\begin{eqnarray}
\mu = \bigoplus_{i} \xi_i, \label{indet_g_1}
\end{eqnarray}
and (\ref{det}) holds again for each $M_i$ on
$\gamma_i \equiv \lambda \oplus \xi_i$:
\begin{eqnarray}
P(M_i = m_{i,k} |\gamma_i) \in \{0,1\},
\ \forall i,\ \forall k, \gamma_i,
\label{indet_g_2}
\end{eqnarray}
this last expression meaning of course that we can write,
again for any of the $M_i$'s,
\begin{eqnarray}
M_i \equiv M_i(\lambda,\xi_i).
\end{eqnarray}

\vspace{0.2cm}\noindent
(b) 
\emph{non $\lambda$-factorisable}, iff
(\ref{indet_g_2}) is not possible for any set of statistically independent
$\xi_i$'s.

\vspace{0.2cm}

We will restrict, for simplicity, our reasonings to just two possible measurements
$A,B \in {\cal M}$, with two possible outcomes, $A,B \in \{+1,-1\}$, all without
loss of generality.
We have seen that, as the more general formulation, we can always write something
like $A = A(\lambda,\xi_A)$, $B = B(\lambda,\xi_B)$.

\vspace{0.2cm}\noindent\emph{Lemma 1}--

\vspace{0.2cm}\noindent
(i) \emph{If $A$ and $B$ are deterministic on $\lambda$},
i.e., (\ref{det}) holds for $A$ and $B$, \emph{then}
they are also $\lambda$-factorisable, i.e.,
\begin{eqnarray}
P(A=a,B=b|\lambda) = P(A=a|\lambda) \cdot P(B=b|\lambda), \label{fac}
\end{eqnarray}
for any $a,b \in \{+1,-1\}$.
Eq. (\ref{fac}) is nothing but the so-called Clauser-Horne factorability
condition \cite{CH74}.

\vspace{0.2cm}\noindent
(ii) \emph{If $A$ and $B$ are indeterministic on $\lambda$}, i.e.,
if (\ref{det}) does not hold for $\lambda$, \emph{then}:
for some $\mu$ (always possible to find \cite{proof_1}) such that now
(\ref{indet_ng}) holds for $\gamma \equiv \lambda \oplus \mu$,
$A,B$ are $\gamma$-factorisable,
\begin{eqnarray}
P(A=a,B=b|\gamma) = P(A=a|\gamma) \cdot P(B=b|\gamma), \label{fac_gamma}
\end{eqnarray}
i.e., (\ref{fac}) holds for $\gamma$, this time not necessarily for $\lambda$.

\vspace{0.2cm}\noindent
(iii) Let (\ref{indet_g_2}) hold for $A,B$, on $\lambda, \xi_A,\xi_B$:
if $\lambda, \xi_A,\xi_B$ are statistically independent,
(hence, $A$ and $B$ are what we have called $\lambda$-factorisable),
then (\ref{fac}) holds for $\lambda$, not necessarily on the contrary.

\vspace{0.2cm}\noindent\emph{Proof}--

\vspace{0.2cm}\noindent
(i) When (\ref{det}) holds, for any $\lambda$ and any
$a,b \in \{+1,-1\}$, $P(A=a|\lambda), P(B=b|\lambda) \in \{0,1\}$,
from where we can, trivially, get to (\ref{fac}).
\hfill\endproof

\vspace{0.2cm}\noindent
(ii) It is also trivial that, if (\ref{indet_ng}) holds, (\ref{fac}) can be
recovered for $\gamma$.
That the same is not necessary for $\lambda$ can be seen with the following
counterexample:
suppose, for instance, that for $\lambda=\lambda_0$, either $A=B=1$ or
$A=B=-1$ with equal probability. It is easy to see that
\begin{eqnarray}
P(A=B=1|\lambda_0) \neq   P(A=1|\lambda_0) \cdot P(B=1|\lambda_0),
\label{counterex}
\nonumber\\
\end{eqnarray}
numerically: $\tfrac{1}{2} \neq \tfrac{1}{4}$.
\hfill\endproof

\vspace{0.2cm}\noindent
(iii) We have, from independence of $\lambda,\xi_A,\xi_B$, and
working with probability densities $\rho$'s:
$\rho_{\lambda}(\lambda,\xi_A,\xi_B) =
\rho_{\lambda}(\lambda) \cdot \rho_{A}(\xi_A) \cdot \rho_{B}(\xi_B)$,
which we can use to write
\begin{eqnarray}
&& P(A=a,B=b|\lambda) = \int P(A=a,B=b|\lambda,\xi_A,\xi_B) \nonumber\\
&&\quad\quad\quad\quad\quad\quad\quad\quad\quad\quad\quad
\times\ \rho_{A}(\xi_A) \cdot \rho_{B}(\xi_B) \ d\xi_A d\xi_B.
\nonumber\\
\end{eqnarray}
and now with the fact that we can recover (\ref{indet_ng}) for $A$ ($B$) on
$\gamma_A = \lambda \oplus \xi_A$ ($\gamma_B = \lambda \oplus \xi_B$),
\begin{eqnarray}
&& P(A=a,B=b|\lambda)
\nonumber\\
&&\quad = \int
P(A=a|\lambda,\xi_A) \cdot P(B=b|\lambda,\xi_B) \nonumber\\
&&\quad\quad\quad\quad\quad\quad\quad\quad\quad\quad\quad
\times\ \rho_{A}(\xi_A) \cdot \rho_{B}(\xi_B)\ d\xi_A d\xi_B
\nonumber\\
&&\quad =
\int P(A=a|\lambda,\xi_A) \cdot \rho_{A}(\xi_A) \ d\xi_A \nonumber\\
&&\quad\quad\quad\quad\quad\quad\quad\quad
\times\ \int P(B=b|\lambda,\xi_B) \cdot \rho_{B}(\xi_B) \ d\xi_B
\nonumber\\
\nonumber\\
&&\quad = P(A=a|\lambda) \cdot P(B=b|\lambda).
\end{eqnarray}
On the other hand, let $\lambda, \xi_A,\xi_B$ be not statistically independent:
we can set for instance, as a particular case, $\xi_i \equiv \mu$, $\forall i$,
therefore reducing our case to that of (\ref{indet_ng}).
Once this is done, our previous counterexample in (ii) is also valid to show that
factorability is not necessary for $\lambda$ here.
\hfill\endproof





\begin{thebibliography}{99}
\bibitem{PDC_explanation}
Parametric Down Conversion (PDC): a pair of entangled photons is obtained by
pumping a laser beam into a non-linear crystal.
\bibitem{note_modes}
For all $t,\mathbf{r}$, one particular mode of the ZPF, with $\mathbf{k}$ the wave-vector
and $\gamma$ the two possible orthogonal polarization states, is always expressible as
\begin{eqnarray}
\mathbf{E}_{0}(\alpha,\alpha^{*}) =
\sum_{\mathbf{k},\gamma}
\left[
\alpha_{\mathbf{k},\gamma}     e^{-i(\omega t - \mathbf{k}\cdot\mathbf{r})} +
\alpha_{\mathbf{k},\gamma}^{*} e^{i(\omega t - \mathbf{k}\cdot\mathbf{r})}
\right],
\end{eqnarray}
where the (stochastic) amplitude $\alpha_{\mathbf{k},\gamma}$ determines, for all space,
a mode $(\mathbf{k},\gamma)$.
Of course, we can always define
\begin{eqnarray}
\alpha_{\mathbf{k},\gamma}(\mathbf{r}) =
\alpha_{\mathbf{k},\gamma}(\mathbf{r}=0) e^{i \mathbf{k}\cdot\mathbf{r}},
\end{eqnarray}
so that now, for a certain mode $\omega$ and at a particular point of space:
\begin{eqnarray}
\mathbf{E}_{0}(\mathbf{r}) =
\sum_{\mathbf{k},\gamma}
\left[
\alpha_{\mathbf{k},\gamma}    (\mathbf{r})  e^{- i\omega t} +
\alpha_{\mathbf{k},\gamma}^{*}(\mathbf{r})  e^{  i\omega t}
\right],
\label{E_0_r}
\end{eqnarray}
where it is obvious that the amplitudes $\alpha_{\mathbf{k},\gamma}(\mathbf{r})$
follow exactly the same probability distribution as $\alpha_{\mathbf{k},\gamma}$,
i.e., their distribution is governed by $W(\alpha,\alpha^{*})$.
This all said and for the sake of simplicity, we shall abuse notation (as is
done customarily in \cite{pdc1,pdc2,pdc3,pdc4,pdc5,pdc6,pdc7}) with
\begin{eqnarray}
\alpha,\alpha^{*} \equiv 
\{ \alpha_{\mathbf{k},\gamma}, \alpha_{\mathbf{k},\gamma}^{*} \}
\end{eqnarray}
i.e., $\alpha,\alpha^{*}$ denotes a set of mode amplitudes (therefore in principle
defined for all space) for a given $\omega$, (one or a set of) relevant wave-vectors
$\mathbf{k}$ and (the relevant) polarizations components $\gamma$.
Finally, of course the overall vacuum field is obtained as a sum of all modes.
\bibitem{note_wvectors}
In the present framework, each wave-vector $\mathbf{k}$ defines a plane wave propagating
through the entire coordinate space, which obliges us to take some precautions.
Let $\{\mathbf{k}_s\}$ and $\{\mathbf{k}_i\}$ be the sets of relevant wavevectors
for the sets of relevant amplitudes $\{\alpha_s\}$ and $\{\alpha_i\}$, respectively,
then we will suppose that $\{\mathbf{k}_s\} \cap \{\mathbf{k}_i\} = \O$;
in the (really very unlikely) case that this did not happen, we can always redefine
$\alpha_i$ so the independence with $\{\alpha_s\}$ still holds.
\bibitem{note_dmodels}
For low intensities, the departure from the quantum mechanical model stands for a
certain ``dead-zone'' and a much higher dark count rate.\\
At first order in perturbation theory, the expectation value of the number of photons
is zero and therefore such rate vanishes: experimentally observed dark counts are
usually associated to thermal effects or electronic noise.
Besides, a saturation effect is introduced, which on the other hand is a feature
of all real detectors, but in this case it appears at much lower intensities
(for energies lower than one photon). 
Consistency with experimental data demands compliance with certain relations provided
in that same paper \cite{W_LHV_02}.
\bibitem{laser_phase}
Indeed, the laser is described by a coherent state, which for a high intensity signal
means that it can be regarded as quasi-classical wave with well defined amplitude and
phase (and introduced as a non-quantized external potential, see \cite{pdc1} for
instance).\\
The phase of this complex amplitude is for instance relevant in expressions (4.10)
from \cite{pdc1}, and has the potential to interfere constructive or destructively
with the $\alpha$'s, increasing or decreasing the overall intensity of the signal,
therefore modifying the detection probabilities.
From this point of view, such phase cannot be at all excluded from the vector of
relevant hidden variables and therefore non-factorability (of detection probabilities)
in $\alpha$ is the most natural feature to expect (though not strictly necessary),
see our detailed analysis in Sec. \ref{Fac}.
\bibitem{crit_eff}
Let us consider a Bell experiment and let $\eta$ be what people in QInf define as
``detection efficiency'' \cite{LHVs} (from our point of view this is clearly
a misleading term, we should call it simply ``detection rate'', whether is
reduced or not by technological imperfections, see Sec. \ref{LR_v_QM}).
A possible Local Hidden Variables (LHV) model for the experiment is then composed by
a set of states $\{s\}$ with probabilistic weights (or frequencies) $\rho_s$:
all restrictions the model must satisfy (either regarding the behavior of detectors
or the quantum mechanical predictions) are linear in $\{\rho_s\}$ and the trivial
solution $\eta=0$ (all $\rho_s=0$ but for the one that predetermines no detections
at all) is always admissible, implying
there is always some $\eta_{\rm crit} \geq 0$ such that for all
$\eta \leq \eta_{\rm crit}$ the desired LHV can be built.
The reasoning applies whether if the LHV just simulates a violation
of a particular Bell inequality or also every other quantum prediction for a given
state and set of observables (the addition of more restrictions would in principle
lower $\eta_{\rm crit}$).
\bibitem{proof_1}
A possible (not unique) procedure to build $\mu$ is this:
for each $M_i$, we define a new random variable $\sigma_i$ and assign
values for each pair $(\lambda,m_{i,k})$: $\sigma_i \equiv \sigma_i(\lambda,m_{i,k})$,
and now simply do
$\mu \equiv \bigoplus_{i} \sigma_i$.
As built, $\sigma_i$'s are not necessarily independent from one another,
nor are they necessarily independent from $\lambda$.
\bibitem{c_Ham}
The PDC model is by construction restricted to a certain subset of QED-states,
obtained directly from a mix of the vacuum state (hence one with positive
Wigner function) with a quasi-classical signal (the laser),
and a time-evolution governed by a quadratic Hamiltonian (hence one that
preserves the positivity of the Wigner function);
for this last point, see \cite{KimNoz} or \cite{Casado_PHD}.
\bibitem{note_WPDC}
For instance, the description of the ``pump'' (laser beam) as a non-quantized,
external potential is just an approximation (one that allows for a Hamiltonian that
is quadratic in creation/annihilation operators, and hence preserves positivity
of the Wigner function);
in any case, we should not forget that further refinements may include not
only quantization of the laser but also additional terms expressing the
interaction with further ZPF modes neglected in the present formulation.
\bibitem{classic_photon}
For instance,
the inclusion of particle sub-structure could also be the way forward
to explain other peculiarities of the quantum world that seem so far
alien to a purely classical framework.
In particular, two very crucial points:\\
(i) the appearance of a discrete spectrum of observable energy exchanges
between matter and the electromagnetic field, due to the presence of
meta-stable states (``attractors'') in the classical phase space of
the system (for instance the Hydrogen atom), once observations are assumed
as some time-average that smears out transients, being therefore reduced
to a discrete spectrum of states and transitions between them;\\
(ii) the possibility of highly directional radiation patterns (emission
of energy) and resonances (absorption of energy), in the same way as they
arise in macroscopic systems with rich spatial structure.\\
Point (i) can potentially accommodate the famous ``quanta'' $E = h\omega$
relation, attempts at which we have left at an intermediate step elsewhere;
point (ii) would make possible the transfer of energy through long
distances with no spread.
\end{thebibliography}
\end{document}